\documentclass[11pt,twoside,A4]{article}
\usepackage{times,fancyhdr}
\usepackage[dvips]{graphicx}
\usepackage{latexsym}
\usepackage[affil-it]{authblk}
\usepackage{mathptm}
\usepackage{amsmath}
\usepackage{amssymb}
\usepackage{floatrow}
\usepackage{units}
\usepackage{setspace}
\usepackage{hyperref}
\usepackage[font=small,labelfont=bf]{caption}
\usepackage[top=4cm, bottom=4cm, left=3.5cm, right=3.5cm]{geometry}

\usepackage{color,graphicx}
\usepackage{caption}
\usepackage{times}
\usepackage{float}
\usepackage[colorinlistoftodos]{todonotes}
\usepackage{marginnote}

\def\beq{\begin{equation}}
\def\eeq{\end{equation}}
\def\beqn{\begin{eqnarray}}
\def\eeqn{\end{eqnarray}}

\begin{document}

\title{A Toy Model for Local and Deterministic Wave-function Collapse}
\author{Sandro Donadi, Sabine Hossenfelder}
\affil{Frankfurt Institute for Advanced Studies\\
Ruth-Moufang-Str. 1,
D-60438 Frankfurt am Main, Germany
}
\date{}
\maketitle
\vspace*{-1cm}

\begin{abstract}
A local, deterministic toy model for quantum mechanics is introduced and discussed. It
is demonstrated that, when averaged
over the hidden variables, the model produces the same predictions as quantum mechanics. In the model considered here, the dynamics depends only on the settings of the measurement device at the detection time, not how those settings were chosen. As a consequence, the model is locally causal but violates statistical independence. We show that it is neither fine-tuned nor allows for superluminal signalling. 
\end{abstract}

\section{Introduction} 
 
A century ago, one might have hoped that quantum mechanics would one day be replaced by a theory compatible with the deterministic and local pre-quantum ontology. No such theory is known today: Neither Bohmian mechanics \cite{Bohm}, nor models with spontaneous collapse \cite{GRW, Bassi:2003gd}, nor stochastic quantum mechanics \cite{Nelson} are local, and the latter two are not deterministic either. 
Bell's theorem \cite{Bell} proves that such a local and deterministic replacement of quantum mechanics -- if it is to reproduce observations -- must violate statistical independence. This option has so-far not been very much explored. 

Our aim here is to put forward a toy model that is local, causal and deterministic, but violates statistical independence, and further the scientific discussion by elucidating its properties. Eventually, the goal of developing such a model is to remove the instantaneous measurement update, and hence make it easier to combine quantum mechanics with general relativity.

Several toy models which violate statistical independence
have previously been proposed \cite{brans,Palmer:1995mxd,degorre,Palmer:2009mxd,Hall3,Sudarsky} (for a review, see \cite{Hall}). The one introduced here has the advantage
that it can be applied to any kind of quantum mechanical system with a finite-dimensional Hilbert space. This new model should not be taken too seriously as a viable description of nature. It is neither pretty nor does it make sense as a fundamental theory for reasons that will be discussed later. However, it will allow us to study the consequences of requiring both locality and determinism.

 We will use units in which $c= \hbar =1$.

\section{The Toy Model}

 Suppose we have a quantum mechanical system described by a state $|\Psi \rangle$ and a Hamiltonian $ H$ in a $N$-dimensional Hilbert space. We want to measure an observable with $N$  eigenstates $| I \rangle$, $I \in \{1,.., N\}$, each one corresponding to a different state of the detector at the time of detection\footnote{Strictly speaking, of course, detection does not happen in a single moment but during a finite time-interval.
This distinction, however, does not matter for the purposes of this toy model as will become clear later.}.
The system could be a composite system and the detector could be distributed over different locations; the following will not depend on these details. If there are multiple detectors, we choose the space-time slicing so that the measurement takes place at the same coordinate time for all detectors.\footnote{If there is no space-like slice that can accommodate all the detectors, then they are causally related, in which case the later detectors can be ignored because the state is already ``collapsed''.}

We then postulate that all the states in the Hilbert-space except the eigenstates of the
measurement observable are unstable under any kind of disturbance that comes from the hidden variables. That is, we will introduce a deviation from quantum mechanics quantified by a parameter, $\kappa>0$, which collapses arbitrary initial states into detector eigenstates. The measurement outcome will then be determined by the initial state of the system and the value of the hidden variables. When we average over these hidden variables, only the dependency on the initial state remains and the probabilistic outcome agrees with that of quantum mechanics. The reader can think of these variables as encoding the detailed
degrees of freedom of the detector because that is the most minimal possibility, but it could be more
complicated.

In the next subsection, we will start with the case $N=2$ to introduce the general idea, and then generalize to larger $N$. 

\subsection{$N=2$}

For $N =2$ we have
\beqn
|\Psi (t) \rangle := a_1(t) | 1 \rangle + a_2 (t) | 2 \rangle ~~,
\eeqn
with complex factors $a_1(t)$ and $a_2(t)$ that fulfill $ |a_1(t)|^2 + |a_2(t)|^2 = 1$. We will denote the time of
system preparation with $t_{\rm p}$ and that of measurement with $t_{\rm d}$. 
Let us further denote with $|\Psi ^* (t) \rangle$ the solution to the normal Hamiltonian evolution
\beqn
|\Psi^* (t) \rangle := \exp ( - {\rm i}  H (t -t_{\rm p}) ) |\Psi (t_{\rm p}) \rangle ~.
\eeqn
It will also be handy to define the coefficients of the eigenvectors at the time of measurement under the
usual Hamiltonian evolution:
\beqn 
\alpha_I := \langle I | \Psi^* (t_{\rm d}) \rangle~.
\eeqn

We know that in quantum mechanics the probability of measuring the $I$-th outcome is $|\alpha_I |^2$, so the
superdeterministic theory must reproduce this on the average. These numbers can be calculated from $ H$
alone, so the underlying dynamics that we are about to construct is unnecessary to make the probabilistic predictions, but this isn't the point. The point of this model is to get rid of the measurement update.

As hidden variables we use complex numbers that are
uniformly distributed inside the complex unit circle. The radius of the distribution could be chosen differently for each of these variables, but this would just add
unnecessary parameters. We will denote the random variable as $\lambda_2$, where the
index $2$ on $\lambda_2$ refers to $N=2$. These hidden variables are part of what fully specifies the initial state. 

With that, we construct, as usual, the density matrix $\rho(t) := |\Psi(t) \rangle \langle \Psi(t) |$ but change the dynamical law to 
\beqn
\partial_t \rho(t) = - {\rm i} \left[  H, \rho(t) \right] +  \kappa \left( L^{\textcolor{white}{\dag}}_{21} \rho(t) L_{21}^\dag - \frac{1}{2} \{ 
\rho_t, L^\dag_{21} L_{21}^{\textcolor{white}{\dag}} \} \right)~. \label{dyn}
\eeqn
Here, $\kappa$ is a constant of dimension energy (for the interpretation, see Section \ref{less}), the curly brackets denote the anti-commutator and
\beqn
L^{\textcolor{white}{\dag}}_{21} &:=& \theta(\sigma_2)  |\chi_2 \rangle \langle \chi_1| + \theta(-\sigma_2)  |\chi_1 \rangle \langle \chi_2| ~, \\
\sigma_2 &:=&  |  \lambda_2 \beta_2 |^2 - (1 -  | \lambda_2|^2) |\beta_1|^2 ~,
\eeqn
where $| \chi_I \rangle$ is an arbitrary orthonormal basis, and 
\beqn 
\beta_I := \langle \chi_I | \Psi^* (t_{\rm d}) \rangle~.
\eeqn
Again, the indices on $L_{21}$ and $\sigma_2$ refer to the case $N=2$ and are merely there to make it easier to later
generalize to higher $N$. 

$\theta(\cdot)$, as usual, denotes the Heaviside-function which we define as $\theta(x) = 0$ for $x \leq 0$ and $\theta(x) =1$ otherwise.
It must be emphasized that the $\rho(t)$ in Eq.\ (\ref{dyn}) is {\it not} the density matrix of quantum mechanics. To get the density matrix of quantum mechanics, we have to average over the random variable $\lambda_2$. We will demonstrate this explicitly later. 

The final ingredient to the model is now the future input: We postulate that the evolution is optimal if and only if $|\chi_I \rangle = |I \rangle$, so that $\alpha_I = \beta_I$. It might seem somewhat perplexing that we introduced the $|\chi_I \rangle$ only to then remove then. We will comment on this later. There are a variety of functions that, when optimized, would spit out this requirement and it is rather unnecessary here to write down a specific one.\footnote{For example $\sum_I (\langle \chi_I|I\rangle + \langle I | \chi_I\rangle)$, when maximized will yield the constraint $|\chi_I\rangle = |I\rangle$. } 

The dynamical law (\ref{dyn}) has the Lindblad-form \cite{Lindblad,Gorini}. It is the master-equation for one of the most common examples of decoherence, that of amplitude damping in a two-level system \cite{Nielsen}. We will in the following work in the limit where $\kappa$ is much larger than the typical enegies of the system, so that the collapse dynamics is the dominant effect. In this regime, it is then easy to explicitly solve (\ref{dyn}) and the solution has the following asymptotic behavior (for details, see Section \ref{fine}):
\beqn
\lim_{\kappa t \to \infty} \rho (t) &=& \begin{pmatrix}
0 & 0 \\
0 & 1
\end{pmatrix} \quad \mbox{for $\sigma_2 > 0$}~, \\
\lim_{\kappa t \to \infty} \rho (t) &=& \begin{pmatrix}
1 & 0 \\
0 & 0
\end{pmatrix} \quad \mbox{for $\sigma_2 < 0$}~. \label{limits2}
\eeqn

\begin{figure*}[h]
\centering
\includegraphics[width=0.5\textwidth]{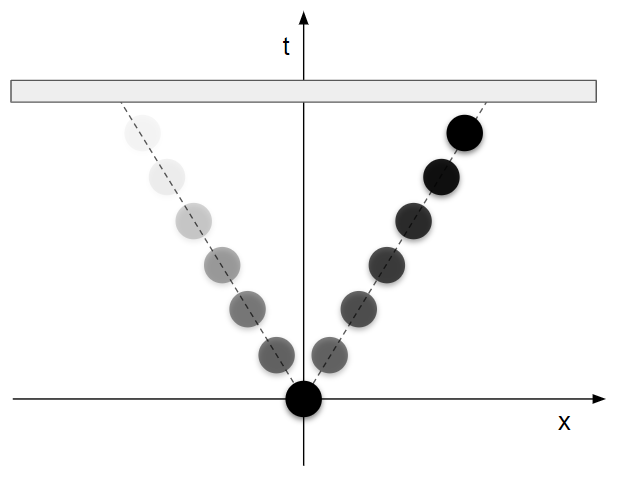}
\caption{Schematic illustration of the collapse described by the toy model. Shades of grey indicade absolute values of amplitudes, where darker (lighter) shades are larger (smaller) amplitudes. Just after the superposition is created (eg, by a beam splitter), the two branches have the same amplitude, but as they approach the detector (grey bar on top), one branch becomes dominant over the other one.}
\label{figcollapse}
\end{figure*}

In Figure \ref{figcollapse} we illustrate the collapse for the simplest case, that of a two-state superposition with equal amplitudes. Contrary to what happens in standard quantum mechanics, the collapse in this model is gradual and it starts before the particle reaches the proximity of the detector. 
This toy model, importantly, is not a model of spontaneous collapse \cite{Bassi:2003gd} or, more generally a model of a stochastic process \cite{stochastic}, because there is no stochasticity in the dynamics. The dynamics is determined by the time-independent random variable and not a random walk. If one selects a specific $\lambda_2$, then one knows for certain what $\sigma_2$ is and one knows what the outcome of the evolution is. 

We now have to show that the probabilities for the state to evolve to $|1 \rangle$ and $|2 \rangle$ come out correctly. As we just saw, in the limit where $\kappa \gg E_{\rm max}$  an initial state $|\Psi(t) \rangle$ will go to $|2 \rangle$ if $\lambda_2 > 0$. We therefore have
\beqn
P( |2 \rangle ) = P (\sigma_2 > 0)~,
\eeqn
where the $P$ here and in the following denotes a probability.
In terms of the random variable $\lambda_2$, we can write the condition $\sigma_2 > 0$ as
\beqn
| \lambda_2 \alpha_2 |^2 &>& (1 -  | \lambda_2|^2)  |\alpha_1|^2  \nonumber\\
\Leftrightarrow \quad \quad \quad  |  \lambda_2 \alpha_2 |^2 &>& (1 -  |\lambda_2|^2) (1- |\alpha_2|^2 ) \nonumber\\
\Leftrightarrow \quad \quad \quad | \lambda_2|^2 &>& 1- |\alpha_2|^2~. \label{prob}
\eeqn
Since $ \lambda_2$ is uniformly distributed in the unit circle, this means we are asking for the area of the ring between the radii $\sqrt{1- |\alpha_2|^2}$  and 1 relative to
the area of the whole disk. The area of the ring and the disk scale with the square of the radius, so the fraction is $1-1+|\alpha_2|^2$ = $|\alpha_2|^2$, and we have
\beqn
P(|2 \rangle) = |\alpha_2|^2~.
\eeqn
 It follows that the probability of the state to evolve to $|1 \rangle$ is $|\alpha_1|^2$. This is exactly Born's rule. 
 
 \subsection{$N>2$}

To generalize from $N=2$ to larger $N$, it helps to have a physical description of the process that the dynamical law (\ref{dyn}) is commonly used to describe. For the case $\sigma_2 > 0$, the state $|\chi_1\rangle$ is unstable and decays to $|\chi_2\rangle$, while for $\sigma_2 < 0$, it is the other way around,  $|\chi_2 \rangle$ decays to $|\chi_1 \rangle$. The constant $\kappa$ determines the decay time. Provided that the contribution from the Hamiltonian can be neglected, this leads asymptotically to the limits in (\ref{limits2}). 

For $N>2$ one then iteratively adds more random vectors $\lambda_N$, each of which is independently uniformly distributed on the unit disk and defines  
\beqn
\sigma_N &:=& | \beta_N  \lambda_N |^2 - \sum_{I=1}^{N-1} |\beta_I|^2 (1 - |\lambda_N|^2 ) ~.
\eeqn

With that, we construct the Lindblad-operators recursively for higher $N$ by drawing on the comparison with decaying states. In the step from $N-1$ to $N$, we introduce $N-1$ new operators 
\beqn
L_{NM} &:=&    \theta(\sigma_N)   |\chi_N \rangle \langle \chi_M| + 
 \theta(-\sigma_N)  |\chi_M \rangle \langle \chi_N| \quad \mbox{for} \quad M \in \{1.. N-1\}~. \label{lnk}
\eeqn
This means that we have in total $N(N-1)/2$ Lindblad-operators $L_{KM}$ for $N$ dimensions, and by convention we have labelled them so that the second index is always strictly smaller than the first. With that, the master equation can be written as
\beqn
\partial_t \rho(t) = - {\rm i} \left[  H, \rho(t) \right] +  
\kappa \sum_{K>M=1}^N  \left( L^{\textcolor{white}{\dag}}_{KM} \rho(t) L_{KM}^\dag - \frac{1}{2} \{ \rho(t), L^\dag_{KM} L_{KM}^{\textcolor{white}{\dag}} \} \right)~. \label{dynN}
\eeqn
As above for the case $N=2$, we obtain the density matrix of quantum mechanics by averaging the $\rho$ from Eq.\ (\ref{dynN}) over the hidden variables.

\begin{figure*}[h]
\centering
\includegraphics[width=7cm]{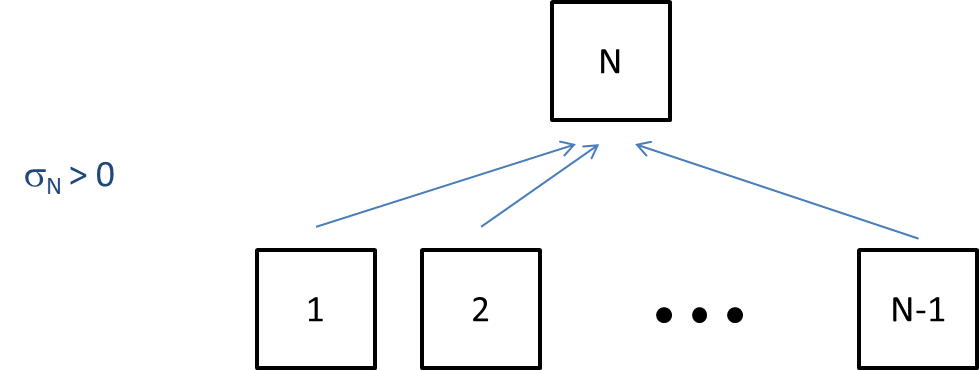}
\includegraphics[width=7cm]{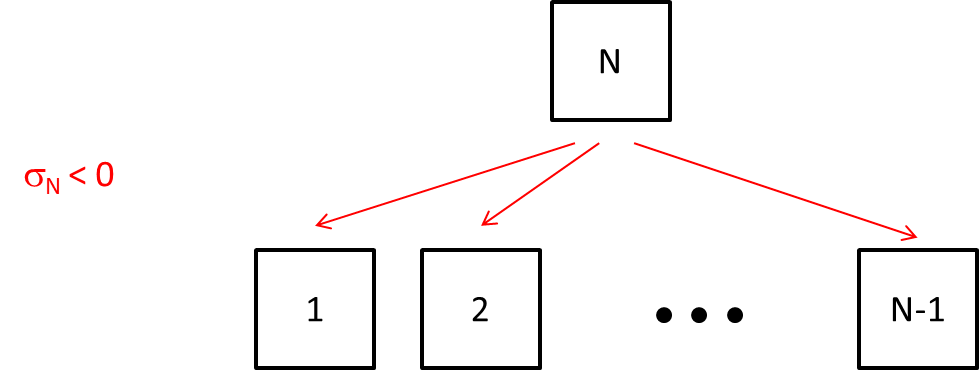}
\caption{The dynamics induced by the Lindblad terms involving the operators $L_{NM}$ with $M=1,2,...N-1$. For $\sigma_N>0$  each state will evolve to the state $|\chi_N\rangle$. For $\sigma_N<0$, the state $|\chi_N\rangle$ will evolve evenly into the states $|\chi_1\rangle,...|\chi_{N-1}\rangle$. \protect{\label{sigma}}}
 
\end{figure*}

Before we move on, let us briefly interpret the structure of the Lindblad-operators that we have constructed. The $N-1$ new operators that we add in each step $N-1$ to $N$ have the property that they either take any initial state $| \chi_K\rangle$ with $K<N$ and asymptotically convert it to $|\chi_N\rangle$ (if $\sigma_N > 0$) or they take the initial state $|\chi_N \rangle$ and distribute it evenly on the states $|\chi_K\rangle$ with $K < N$ (if $\sigma_N < 0$). This is illustrated in in Figure \ref{sigma}.

So, if $\sigma_N>0$, we know that the result will be $|\chi_N\rangle$. Let us then look at the case $\sigma_N \leq 0.$ Once the state $|\chi_N\rangle$ is cleared out and distributed over the subspace spanned by $|\chi_1\rangle ... |\chi_{N-1}\rangle$, the same thing happens again for the remaining states, but this time depending on the sign of $\sigma_{N-1}$: If $\sigma_{N-1} >0$, then all remaining states go to $|\chi_{N-1}\rangle$, if $\sigma_{N-1} < 0$, then they will be distributed over the subspace spanned by $|\chi_1\rangle ... |\chi_{N-2}\rangle$, and so on. 

Having said that, we can now show that the generalization to $N>2$ still fulfills Born's rule. First, we postulate as previously that the optimal evolution is the one with $|\chi_I \rangle = |I \rangle$, so let us then look at the eigenstate $|N\rangle$. As just explained, if $\sigma_{N} > 0$, then the operators $L_{NK}$ will convert any initial state $|K\rangle$ with $K<N$ into the state $|N\rangle$. One can calculate the probability for this the same way as above in (\ref{prob}) and easily finds
\beqn
P(|N\rangle) = P(\sigma_N > 0) = |\alpha_N|^2~.
\eeqn
Correspondingly, the probability that the state goes to any one of the other $N-1$ eigenstates is $1-|\alpha_{N}|^2$. 

To obtain the probability that an initial state asymptotically goes to $|N-1\rangle$, one then needs to calculate the probability for
\beqn
P(|N -1\rangle) = P(\sigma_N < 0 \wedge \sigma_{N-1} > 0) = P(\sigma_N < 0) P(\sigma_{N-1} > 0)~,
\eeqn
because the random variables are by assumption uncorrelated. 

For this, one has to take into account that for $N-1$ basis states  $\sum_{i=1}^{N-1} |\alpha_i|^2 \leq 1$, and therefore the probability for $\sigma_{N-1}$ to be larger or smaller than one has to be divided by this sum. One sees this by rewriting the requirement $\sigma_{N-1} > 0$ as
\beqn
|  \lambda_{N-1} \alpha_{N-1} |^2 &>& (1 -  |\lambda_{N-1}|^2)  \sum_{I=1}^{N-2} |\alpha_I|^2  \nonumber\\
\Leftrightarrow \quad \quad \quad |  \lambda_{N-1}|^2 \frac{|\alpha_{N-1} |^2}{\sum_{I=1}^{N-1} |\alpha_I|^2}  &>& (1 -  | \lambda_{N-1}|^2) \left(1- \frac{|\alpha_{N-1}|^2}{\sum_{I=1}^{N-1} |\alpha_I|^2} \right) \nonumber\\
\Leftrightarrow \quad \quad \quad| \lambda_{N-1}|^2 &>& 1- \frac{|\alpha_{N-1}|^2}{\sum_{I=1}^{N-1} |\alpha_I|^2}~. \label{prop2}
\eeqn

So, by the same argument as before, the probability for $\sigma_{N-1} > 0$ is now $|\alpha_{N-1}|^2/\sum_{i=1}^{N-1} |\alpha_i|^2$ (instead of just $|\alpha_{N-1}|^2$, as it was for $N-1$ states). Hence we have
\beqn
P(|N-1 \rangle) = ( 1- |\alpha_{N}|^2)  \frac{ |\alpha_{N-1}|^2}{\sum_{i=1}^{N-1} |\alpha_i|^2} = |\alpha_{N-1}|^2~,
\eeqn
which is what Born's rule requires. 

It works the same way for all remaining eigenstates. For example, the probability of an initial state to go to $|N-2 \rangle$ is given by
\beqn
P(|N-2\rangle) &=& P(\sigma_{N} < 0 \wedge \lambda_{N-1} < 0 \wedge \sigma_{N-2} >0) \nonumber \\
&=&  
P(\sigma_{N} < 0) P(\sigma_{N-1} < 0)P (\sigma_{N-2} >0)~,
\eeqn
 which is
\beqn
( 1- |\alpha_{N}|^2)  \left( 1- \frac{ |\alpha_{N-1}|^2}{\sum_{i=1}^{N-1} |\alpha_i|^2} \right) \frac{ |\alpha_{N-2}|^2}{\sum_{i=1}^{N-2} |\alpha_i|^2} = |\alpha_{N-2}|^2~.
\eeqn
In general, then, the probability for any initial state to asymptotically come out as $|K\rangle$ is 
\beqn
P(| K \rangle)  &=& P(\sigma_{N} < 0 \wedge...\wedge \sigma_{K+1} < 0 \wedge \sigma_{K} >0) \nonumber \\
&=& (1- |\alpha_{N}|^2) \prod_{J=N-1}^{K+1} \left( 1- \frac{ |\alpha_{J}|^2}{\sum_{I=1}^{J} |\alpha_I|^2} \right) \left(\frac{ |\alpha_{K}|^2}{\sum_{I=1}^{K} |\alpha_I|^2}\right) = |\alpha_{K}|^2~.
\eeqn

This completes our proof that Born's rule is fulfilled for all $N$.

The way we have defined the $N$-dimensional case depends on the order of the basis-states. Since the order is arbitrary, this seems unphysical. It does not really matter for the purposes of this model because the resulting probability distribution is independent of the ordering. However, one could make the model explicitly independent of the ordering by summing over all possible permutations. 

One may note that this distribution fulfils the requirements for Born's rule laid out in \cite{Hossenfelder:2020gdb}. 

\section{On future input dependence}

\begin{figure*}[h]
\centering
\includegraphics[width=0.9\textwidth]{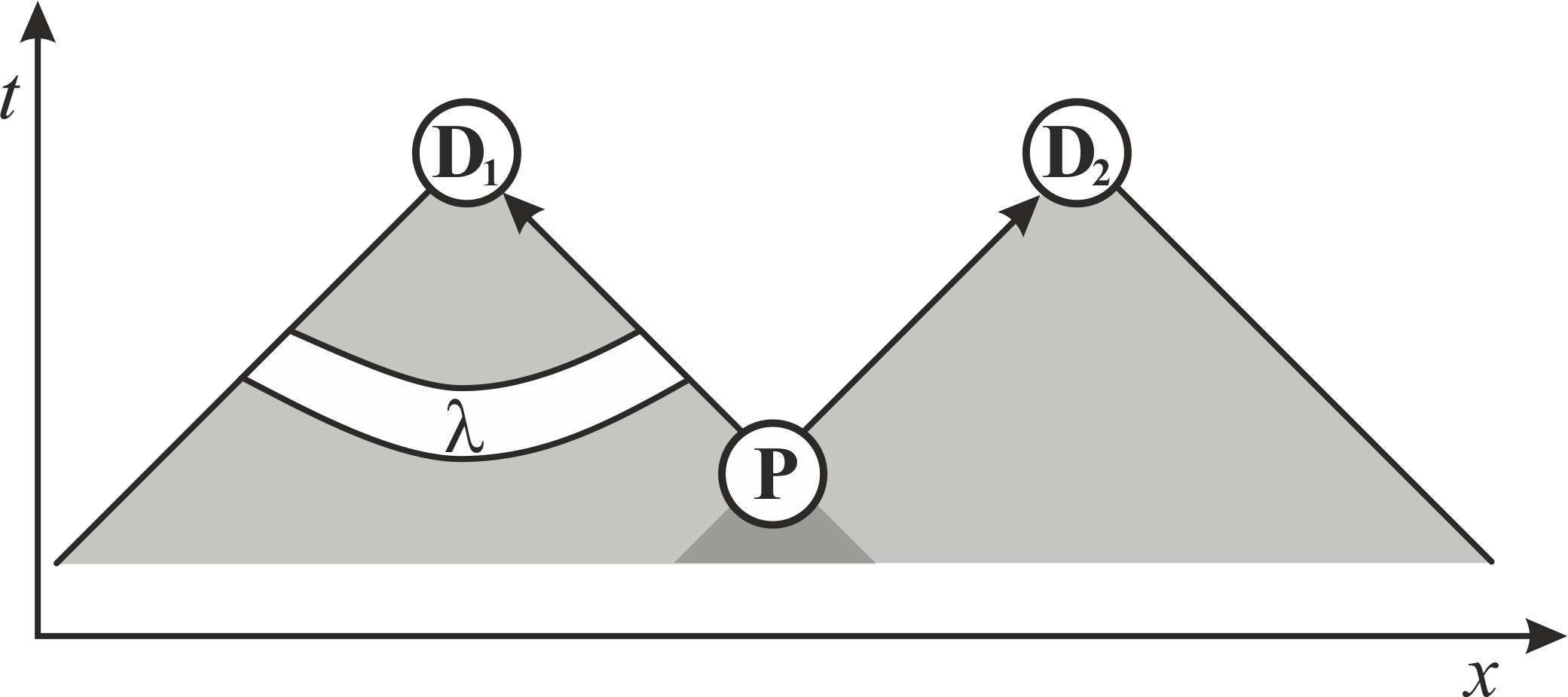}
\caption{Sketch to illustrate local causality.}
\label{bell}
\end{figure*}

The key property of the model discussed here is that the evolution of the prepared state depends on what the measurement setting will be at the time of measurement; it has hence a ``future input dependence''  \cite{Wharton}. It is easy to see how this can restore locality and determinism if we have a look at the sketch for Bell's notion of local causality (Figure  \ref{bell}). Here, the {\bf P} indicates the location of preparation, and ${\bf D}_1$ and ${\bf D}_2$ are the two detectors in a typical Bell-type experiment. 

Local causality, in a nutshell, says that if all variables are ``fully specified'' in the backward lightcone of ${\bf D}_1$, then whatever happens at ${\bf D}_2$ provides no further information (for a discussion of this point, please see \cite{Norsen2011}). Quantum mechanics violates this condition because the setting of the second detector matters for the outcome at the first detector. The model we discussed here has the information about both detector settings at ${\bf P}$. This means also that the information about the detector setting ${\bf D}_2$ is already in the backward lightcone of ${\bf D}_1$ (and vice versa). Hence, whatever happens at ${\bf D}_2$ provides no further information.  Note that this is a local requirement because the place of preparation is necessarily in causal contact with both detectors. 

Because the full specification of the hidden variables includes information about the detector settings, this model violates statistical independence. Models of this type are said to be superdeterministic. The purpose of the the toy model introduced above is to demonstrate that for such a theories to reproduce quantum mechanics, one does not need to finetune any input. 

An often-raised question is how the hidden variables at ${\bf P}$ already ``know'' the future detector settings at ${\bf D}_1$ and ${\bf D}_2$. As for any scientific theory, we make assumptions to explain data. The assumptions of a theory are never explained within the theory itself (if they were, we wouldn't need them). Their scientific justification is that they allow one to make correct predictions. The question how the hidden variables ``know'' something about the detector setting makes equally little sense as asking in quantum mechanics how one particle in a singlet state ``knows'' what happens to the other particle. In both cases that's just how the theory works. The difference is that in our toy model the required correlations are within the lightcone, hence compatible with general relativity.

As discussed previously in \cite{otherpaper}, there are two different notions of causality. In the foundations of physics, the widely used one is space-time causality. If one uses space-time causality, the event in the past is by definition the cause. This notion may or may not agree with the definition of causality used in causal models which we may refer to as interference causality \cite{Spirtes,Pearl}. In our toy model, these two notions of causality do not agree, which is why the model is future input dependent, and also the reason why it is not finetuned.

\subsection{Future input dependence and optimization principle: a classical example}

We here want to offer a classical example for future input dependence to illustrate how it can come about by an optimization principle. 

A familiar example for an optimization principle is the principle of least action: Given a Lagrangian for a particle, as well as an initial and final position, we can calculate the trajectory by minimazing the action.  In this case we can always express the evolution law as an Euler-Lagrange equation.
This is not the type of optimization we will will consider here. We are interested instead in the case where we also allow the final point to vary. That is, we evaluate the action for all possible paths for all possible endpoints. 

For a quantum system, the optimal evolution should then be one in which the prepared state always ends up in one of the possible detector eigenstates. Note that this is {\sl not} what happens in quantum mechanics. In quantum mechanics the state (generically) ends up in a superposition of detector eigenstates. The wave-function collapse then needs to be added as an ad-hoc prescription because the Schr\"odinger equation alone fails to give the result that we actually observe. To replace the collapse postulate with a principle of least action, one would need to know the right action to vary over. This is beyond the scope of this present paper. However, we can think of the toy model discussed in the previous section -- which has the detector settings in the evolution law -- as an effective description of an optimization over a final state. 

\begin{figure*}[h]
\centering
\includegraphics[width=0.5\textwidth]{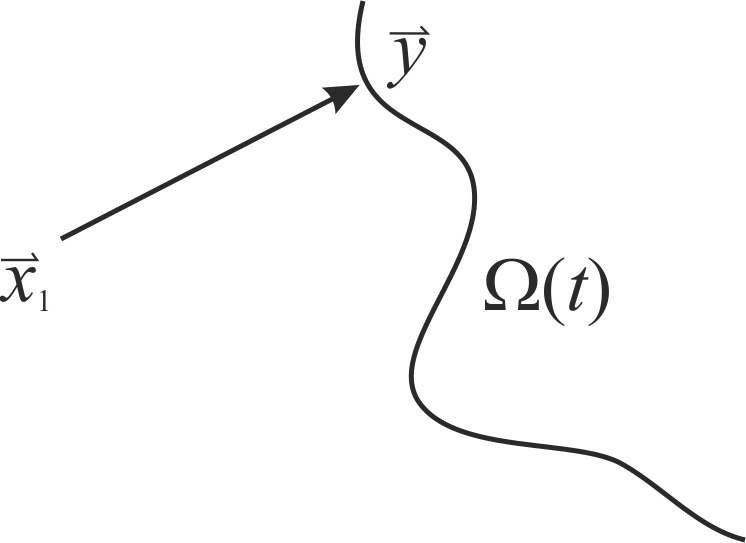}
\caption{Classical example for optimization problem that cannot be expressed by Euler-Lagrange equations.}
\label{toy02}
\end{figure*}

To see how an optimization principle results in future-input dependence, consider the following classical optimization problem. A particle with location $\vec{x}(t)$ moves in 3-dimensional space in absence of a potential, so we have the Lagrangian ${\cal L} := \dot{\vec{x}}^2/2$ and action $S:= \int {\rm d}t {\cal{L}}(\vec{x},\dot{\vec x})$ where a dot denotes a time-derivative, d/d$t$. In the usual principle of least action, we consider a start-time, $t_1$, and an end-time, $t_2$, and fix the start-point and end-point to $\vec{x}_1:=\vec{x}(t_1)$ and $\vec{x}_2:=\vec{x}(t_2)$. Then we can derive the Euler-Lagrange equations, which in this case come out to be simply $\ddot{\vec{x}}=0$.

Now, however, we want to look at a different optimization problem. We start at time $t_1$ at place $x_1$, but at time $t_2$ the particle's location can be any point $\vec{y} \in \Omega$, where $\Omega$ is a 2-dimensional surface (see Figure \ref{toy02}). That is, in addition to the path itself, we also allow the end-point to vary. What is the path of least action now?  In this case the derivation of the Euler-Lagrange equations cannot be {directly} applied. We can however work around this. 

For this, we first keep the end-point fixed, as usual, and then integrate the Euler-Lagrange equation once. This gives $\dot{\vec{x}} = (\vec{y}-\vec{x}_1 )/\Delta t$, where $\Delta t = t_2 - t_1$. We can insert this into the Lagrangian and integrate it to obtain the action as a function of $\vec{y}$. We can then vary the action again, this time over $\vec{y}$. The result is that the optimal path is the one to the point on $\Omega$ that is closest to $\vec{x}_1$. This is because we have kept the time-difference fixed, so a shorter distance means a smaller velocity, which means a smaller action. 

Let us denote this nearest point on $\Omega$ with $\vec{y}_{\rm n}$. We can then insert this into the once integrated Euler-Lagrange equation which becomes $\dot{x} = (\vec{y}_{\rm n} -\vec{x}_1 )/\Delta t$. The final step is then to allow that $\Omega$ could be a function of time, $\Omega(t)$, in which case the nearest point will also be a function of time $\vec{y}_{\rm n}(t)$. The point that minimizes the action is then the nearest point at time $t_2$. We finally get $\dot{x} = (\vec{y}_{\rm n}(t_2) -\vec{x}_1 )/\Delta t$. 

Of course it is utterly trivial to integrate this equation, but deriving this equation wasn't the point. The point was to show that it is possible to rewrite an optimization principle in which the end-point is also varied into a differential equation. This equation will depend on information about the optimization constraint {\sl in the future}, which we do not have. This does not mean that one cannot make a prediction. Even if we do not know the time-dependence of 
$\Omega(t)$, we can very well predict that when we detect the particle, it will be at the point that was nearest to $x_1$. 

This example is just meant to illustrate the general idea for why there is future input in the evolution law of the toy model. Loosely speaking, the surface $\Omega(t)$ is an analogy to the detector setting. The most important difference between the cases is that in quantum mechanics one cannot make a measurement before the final time, because the final time is by definition the time of measurement. 

We want to stress that one does not need to know this future input to make predictions with the model. One simply solves the equation for any possible future input, and makes a conditional prediction. (If input is this, then the prediction is that.) We do the same in quantum mechanics: When we make a calculation in normal quantum mechanics, we don't know what the measurement setting will be. We simply calculate what happens for any possible measurement setting. This is also what we do in our toy model. 

The important difference between the toy model and quantum mechanics is that in quantum mechanics the time evolution of the state independ of the measurement settings. The settings only enter in the calculation of probabilities at the time of measurement. This is the reason why one needs a non-local update of the wave-function and why a realist interpretation of the wave-function is difficult to reconcile with relaitivy. In our future input dependent toy model, on the other hand, the settings determine the local evolution, which avoids a non-local update upon measurements. 

\section{Properties of the toy model}

\subsection{Superdeterminism}

It might seem perplexing why we didn't just put the measurement basis into the evolution law right away\footnote{Indeed, we did this in the first version of this paper.}. The reason is that this way it is easier to see what are the ``hidden variables'' in Bell's sense. Since we already used $\lambda$, we will denote Bell's hidden variables with $\Lambda$. To fully specify the evolution of the prepared state, we need both the $\lambda_I$ and the basis $|\chi_I\rangle$. 
Hence Bell's hidden variables are the combination of both $\Lambda = (\lambda_I, | \chi_I \rangle)$. This makes clear that once we assume $|\chi_I\rangle = |I\rangle$, the hidden variables are $\Lambda = (\lambda_I, | I \rangle)$.

If one does not first introduce the $|\chi_I \rangle$ basis, it is easy to confuse oneself over what the hidden variables are, because Bell lists the detector settings separately. Introducing the $|\chi_I \rangle$ makes clear that the detector settings are part of the full specification of beables that  determines the measurement outcome.  Since 
the variables $\sigma_I$ are functions of the $\lambda_I$ and $|I\rangle$, one can alternatively write $\Lambda = (\sigma_I, |I \rangle)$.

This means that if we consider a Bell-type test with two detectors that have settings {\bf a} and {\bf b}, then the condition
\beqn
p(\Lambda | \mathrm{\bf a}, \mathrm{\bf b}) = p(\Lambda) 
\eeqn
is not fulfilled. Such theories are known as superdeterministic \cite{Hossenfelder:2019shy}. They are the only known consistent, local, and deterministic completion of quantum mechanics. If one mistakes the $\lambda_I$ alone for Bell's hidden variables, then one erroneously arrives at the conclusion that the model respects statistical independence (because the $\lambda$s do not depend on the detector settings) but violates local causality (since one needs the detector settings to calculate the probabilities).

\subsection{Locality}

Let us denote the measurement outcome at the detector with setting {\bf a} as $O_{\rm \bf a}$ and that at the detector with setting {\bf b} as $O_{\rm \bf b}$. 
Then, Bell's notion of local causality is the condition that
\beqn
p(O_{\rm \bf a},O_{\rm \bf b}| {\mathrm {\bf a}}, {\mathrm {\bf b}}, \Lambda)  = p(O_{\bf \rm a}| {\mathrm {\bf a}}, \Lambda) p(O_{\bf \rm b}| {\mathrm {\bf b}}, \Lambda) ~.
\eeqn
This condition is fulfilled because in our toy model the distribution for $\Lambda$ depends on {\bf a} and {\bf b}, so the settings are redundant for the calculation of the probability. Indeed for the toy model we can write:
\beqn
p(O_{\rm \bf a},O_{\rm \bf b}| {\mathrm {\bf a}}, {\mathrm {\bf b}}, \Lambda)  = p(O_{\bf \rm a}| \Lambda) p(O_{\bf \rm b}| \Lambda) ~.
\eeqn

\subsection{Finetuning, Absence Of}
\label{less}

Theories that violate statistical independence are usually dismissed as ``conspiracy theories'' that requires ``fine-tuning''. Those arguments, which originated in the 1970s \cite{SC,Bell77} were previously addressed in \cite{Hossenfelder:2019shy,otherpaper}, and our toy model demonstrates just exactly why these fine-tuning arguments are misleading.

What we mean by ``fine-tuning'' here is that the model requires one to specify so many details to make a prediction that it becomes scientifically useless. It is already clear from the above that the toy model is not fine-tuned: It gives us exactly the same predictions as quantum mechanics without the need to specify many details. But just why are the fine-tuning arguments wrong?

These arguments have it that there are many different ways to choose detector settings, and the model must work for
all of them. This, it is then argued, requires that the model specifies what happens for a huge number of different cases and must still
reproduce quantum mechanics on the average. 

The problem with this argument is that there is no reason why, in a model that violates statistical independence, the measurement outcome should depend on
how the detector settings were chosen any more than this is the case in quantum mechanics. The settings at the time of measurement alone 
are enough to get the correct average predictions. What happens in the brain of the experimenter (or with any other mechanism used to determine detector settings \cite{Leung:2017ndn,bigbell,Friedman:2018byq}) does not matter; the only thing that matters is the setting of the measurement device at the time of measurement. It follows that the average outcome is insensitive to the details, hence not fine-tuned. 

We already know this is possible because it's what we do in quantum mechanics itself: We predict average measurement outcomes from the 
detector settings alone.

A few words are in order here about the notion of fine-tuning that we are using. The expression `fine-tuning' is used in many different areas of physics, but with slightly different meanings \cite{Hossenfelder:2018ikr,lim}. Most of these notions of fine-tuning are meta-physical, that is, they are assumptions which physicists make that have no empirical support, but are merely used to select one model out of many possible models.  That an assumption is meta-physical does not make it wrong, but it means it is optional. Such assumptions are in many cases used out of tradition, not because of empirical necessity.

Meta-physical fine-tuning arguments typically analyze the behavior of the predictions of a model under a variation of input parameters which are not known to be variable. This is the case, for example, in cosmology when constants of nature are varied, or in particle physics, when perturbations in variables (which are physically possible) are conflated with perturbations in parameters that cannot change in any physically possible process. It is also the case in the arguments put forward in \cite{finetuning,Valentini1,Valentini2}. 

These analyses -- while technically entirely correct -- are based on assumptions about probability measures of input parameters which are empirically unknown (or even, qua assumption, unknowable). The output of a model might very well sensitively depend on some input parameter, yet if that input parameter cannot vary, or its variations are constrained, this dependence is irrelevant for the predictability of the model. 

This is also the case for initial values, which are yet another type of input for a model. 
Unless there is empirical evidence that an initial value for a variable can indeed take on different values with a certain frequency, assumptions about the distribution of such initial values are meta-physical. This is especially true for initial values for the state of the universe for which we can never empirically determine the distribution over initial values: Our universe had one set of initial values and that's the one and only instance we can observe. 

Meta-physical arguments are  subject of philosophy. This does not mean they are irrelevant per se, but since meta-physical fine-tuning arguments about superdeterminism were previously addressed in \cite{Hossenfelder:2019shy,otherpaper}, they do not concern us here.

In our analysis, in contrast, we take a strictly instrumental approach. A model is not scientifically viable if it is impossible to make predictions with it, or if making predictions with it requires more information as input than the model delivers as output. Such a model, that requires one to specify a lot of detail to make a prediction, is what we call `fine-tuned'. This, is \emph{not} the definition of fine-tuning used in \cite{finetuning,Valentini1,Valentini2}. This is because the definition used in these other works does not tell us what it actually takes to make predictions with a model. The notion of fine-tuning we use here is the notion that underlies the criticism first raised in \cite{SC,Bell77} and later repeated many times (for more references see \cite{Hossenfelder:2019shy,Chen:2020yoa}), according to which superdeterminism is supposedly unscientific.  

It would be possible to quantify the predictability of our model, but this is both unnecessary and uninstructive because any such quantifier would be arbitrary anyway. We have instead compared our model to standard quantum mechanics and demonstrated that they come out in a tie. Our toy model, hence, is not any more or less fine-tuned than ordinary quantum mechanics.

\subsection{Superluminal Signalling, Absence Of}
\label{fine}

It has been argued in the literature that superdeterministic models require fine-tuning to prevent that entanglement can be exploited for superluminal signalling \cite{finetuning,Bendersky}. Given that there is no finetuning in our model, we will therefore now have a close look at this issue.

We could simply argue that since the toy model was constructed to reproduce the predictions of quantum mechanics, 
faster than light signalling is impossible with the toy model just because it is impossible with standard quantum mechanics. 
One may further notice that the master equation (\ref{dynN}) is linear in the density matrix: the model hence does not suffer from the problem discussed in \cite{Gisin:1989sx,Bassi:2015jka} \footnote{It should be mentioned that \cite{Gisin:1989sx,Bassi:2015jka} were concerned with a different question than the one investigated here, namely whether non-linear generalizations of the Schr\"odinger-equatifon allow for superluminal signalling. Their answer is ``Yes'' for deterministic evolutions and ``in some cases'' for stochastic dynamics. But the equation we use here is neither non-linear nor is it a Schr\"odinger-equation. 
}. However, while correct, these arguments are not very illuminating. Just exactly how does it happen that superluminal signalling is prevented even though the model is deterministic? The calculation in this section will demonstrate explicitly how the average over the hidden variables removes the possibility for superluminal signalling. It will also be instructive to see how the model works in practice.

We will concretely look at a singlet state, shared between two -- potentially very distant -- observers, Alice ($A$) and Bob ($B$),
who measure the spin of a particle, which is either up ($\uparrow$) or down ($\downarrow$) in their basis. We will chose the weights of the entangled state to be $q$ and $\sqrt{1-q^2}$, respectively, (instead of using $1/\sqrt{2}$ for both) to make sure cancellations between terms do not occur because of the symmetry of the state.

The total dimension of the Hilbert space is $N=2^2 = 4$. We now assume the optimization condition is fulfilled $|\chi_I\rangle=|I\rangle=$ and, to connect with the notation of the previous section, we define:
\beqn
 |1 \rangle := |\downarrow_{A} \rangle \otimes |\uparrow_{B} \rangle \equiv |\downarrow_A \uparrow_B \rangle ~&,&~
 |2 \rangle := |\uparrow_{A} \rangle \otimes |\downarrow_{B} \rangle \equiv |\uparrow_A \downarrow_B \rangle ~,\nonumber \\
|3 \rangle := |\downarrow_{A} \rangle \otimes |\downarrow_{B} \rangle \equiv |\downarrow_A \downarrow_B \rangle 
~&,&~ |4 \rangle := |\uparrow_{A} \rangle \otimes |\uparrow_{B} \rangle = |\uparrow_A \uparrow_B \rangle ~.
\eeqn

The prepared state at time $t=t_{\rm p}$ is
\begin{equation}
|\Psi (t_{\rm p}) \rangle= q |\uparrow_{A}\downarrow_{B}\rangle- \sqrt{1-q^2}|\downarrow_{A}\uparrow_{B}\rangle = q |1\rangle- \sqrt{1-q^2}|2\rangle~,
\end{equation}
and has the density matrix
\beqn
\rho(t_{\rm p}) = q^2  |1\rangle \langle1| + (1-q^2)|2\rangle \langle 2| - q \sqrt{1-q^2} \left( |1\rangle \langle2| +  |2\rangle \langle1| \right) ~.  \label{rhoinitial}
\eeqn
Our task is now to solve the master-equation (\ref{dynN}) for the density matrix with the initial value (\ref{rhoinitial}) and then take the average over the hidden variables to get
\beqn\label{roqm}
\overline \rho(t) := \int \rho(t) \prod_{I=2}^4  d \lambda_I ~.
\eeqn

For a 4-dimensional Hilbert-space, we have 6 Lindblad-operators, one for each entry above the diagonal. For the singlet state under consideration, $\alpha^2_1 =q^2, \alpha_2 = 1-q^2, \alpha_3^2 = \alpha_4^2 = 0$, so we have
\beqn
\sigma_2 = |\lambda_2|^2 - q^2 ~,~ \sigma_3 = |\lambda_3|^2 -1 ~,~\sigma_4 =  |\lambda_4|^2 -1~.
\eeqn
Since the $\lambda$'s all lie within the unit circle, we see that $\sigma_3$ and $\sigma_4$ are both $\leq 0$. This reflects the fact that the probability to end up in the state $|3\rangle$ or $|4 \rangle$ should be zero. 

One can then integrate the master-equation (see Appendix), using the initial state (\ref{rhoinitial}), from which one obtains  the evolution for the matrix elements $\rho_{ij}(t):=\langle i|\rho(t)|j\rangle$
\beqn
\rho_{11}(t) &=& \theta(\sigma_2) q^2 e^{- \kappa t}  + \theta(-\sigma_2) \left(1 - \left(1-q^2 \right)e^{ -\kappa t}  \right) ~,~\nonumber \\
 \rho_{22}(t) &=& \theta(\sigma_2) (1-  q^2 e^{- \kappa t}) + \theta(-\sigma_2) \left( 1 - q^2 \right) e^{ -\kappa t}
~,\nonumber \\
\rho_{12}(t) =\rho_{22}(t) &=& q\sqrt{1-q^2} e^{- \kappa t/2} ~. \label{solsN4}
\eeqn
and all the other matrix elements are equal to zero. Note that  ${\rm Tr}(\rho) = 1$ for all times. 
 
We already know that
\beqn
P(\sigma_2 < 0) = q^2~,~P(\sigma_2 > 0) = 1-q^2~. \label{probs}
\eeqn
So for the average of the density matrix 
we have
\beqn
\overline \rho_{11}(t) &=& (1-q^2) q^2 e^{- \kappa t}  + q^2 \left(1 - \left(1-q^2 \right)e^{ - \kappa t}  \right) = q^2 ~, \nonumber \\
\overline \rho_{22}(t) &=& (1-q^2)(1-  q^2 e^{- \kappa t})  + q^2\left( 1 - q^2 \right) e^{ -\kappa t} = 1- q^2 ~, \nonumber \\
\overline \rho_{12}(t) = \overline \rho_{22}(t) &=& q\sqrt{1-q^2} e^{- \kappa t/2} ~. \label{fin}
\eeqn
This means the only thing the Lindblad-operators do is to exponentially suppress the off-diagonal terms in the density matrix. 
We can now take the partial traces
\beqn
\overline \rho_A &=& {\rm Tr}_B (\overline \rho) = q^2  |\uparrow_{A} \rangle \langle  \uparrow_{A} | + (1-q^2) |\downarrow_{A} \rangle \langle  \downarrow_{A} | ~,\\
\overline \rho_B &=& {\rm Tr}_A (\overline \rho) = q^2  |\downarrow_{B} \rangle \langle  \downarrow_{B} | + (1-q^2) |\uparrow_{B} \rangle \langle  \uparrow_{B} | ~.
\eeqn
This is exactly the same result as in quantum mechanics.
We thus see that if one treats the random variables in the above discussed toy model as truly random then superluminal signalling is not possible. 

However, one may object that if the hidden variables are really random, then the model is not superdeterministic because it is not deterministic in the first place. Indeed, in a superdeterministic model, the randomness should not be fundamental but merely a consequence of lacking detailed information -- as one expects in a `hidden variables' theory. Thus, the distribution of the hidden variables should in certain circumstances deviate from being completely uniform. In this case, taking the average over the hidden varables will not give the same result as in quantum mechanics. This is apparent from Eqs.\ (\ref{solsN4}) which, when averaged with probabilities other than (\ref{probs}) will not give (\ref{fin}), but have correction terms in whatever is the deviation of the distribution of the $\lambda$'s from uniformity.

This per se does not constitute an opportunity for signalling, because to exploit this channel one of the observers would have to be able to skew the distribution of hidden variables so that the measurement outcome for the other observer would be correlated with information they desire to send. What it takes to do that is, again, a question this toy model cannot address because it does not specify the origin of the randomness. However,  that so far no experiment has revealed any deviations from Born's rule suggests it is not easy to prepare a system in a state where the distribution of hidden variables does not reproduce quantum mechanics. 

Nevertheless, one can certainly speculate about different distributions of the random variables. This has been done for example for pilot-wave theory \cite{Val1,Val2,Val3}, where deviations from quantum equilibrium might be observable in astrophysical
and cosmological tests \cite{Val4,Val5}.

\section{Discussion}

One should not think of this model as a viable description of nature because the way that the random variables enter the dynamics has no good motivation. It is a toy model for an effective limit of a more fundamental theory. The parameter $\kappa$, for example, defines a time-scale like for decoherence and should be understood as induced by the interaction with a larger system that has been integrated out. That is, $\kappa$ scales with the total size of the system and in the limit when the prepared state has no further interaction, $\kappa = 0$, one just has normal quantum mechanics.

This is also why the model is not Lorenz-invariant: The environment defines a preferred frame, yet the environment does not explicitly appear. The model further violates energy conservation. Again, this is because it stands in for an effective description that, among other things, does not take into account the recoil (and resulting entanglement with) parts of the experimental equipment. For the same reason, as mentioned in footnoote 2, the time of measurement is not really {\emph{the}} time of measurement, but a time at which we would for all practical purposes say the measurement has been completed. 

Furthermore, since the operators $L_{NK}$ are defined in terms of the eigenstates of the observable measured at the time of detection, we have still used an external definition of detection that is not included in the definition of the model. Now, the pointer basis defining these operators can be identified using the einselection methods introduced by Zurek \cite{Zurek} when one includes an environment, so this is not a problem per so. However, a more fundamental model is needed to explain how these operators come to couple to the prepared state the way that they do. 

This toy model avoids non-local interactions by hard-coding the dependence on the detector settings into the evolution law.  This is another reasons why one should not take this model too seriously: A good, fundamental, model should allow us to {\sl derive} that the effective law for the prepared state depends on the detector settings. This requires that the to-be-found fundamental model includes the detectors and the environment and possibly other transformation devices that are part of the experimental setup. This has to be the case because otherwise we would lack information to define what the detector eigenstates are.

All these issues are resolvable in principle, but given that this model is not intended to make a lot of sense, putting more effort into it seems not a good time-investment. The reason for this little exercise was merely to demonstrate that there is nothing to fine-tune here. 
For a Hilbert-space of dimension $N$, we have $N-1$ uniformly distributed, complex, random variables. Picking a set of specific values for these variables will in the limit $\kappa t \to \infty$ lead to one particular detector eigenstate: No collapse postulate necessary. But if we do not know what the values of the random variables are, we average over all possible values and get Born's rule. The model has one free parameter, $\kappa$, but its value doesn't matter as long as it is much larger than the typical energies of the system, so that the collapse dynamics is dominant. One may not like this construction for one reason or the other, but clearly there is no huge number of details to be specified here.

This works regardless of what the Hamiltonian is, how large the system is, how many detectors there are, or what observable is being measured. It will work for the double-slit\footnote{Double slit experiments are typically described using infinite Hilbert-spaces, which we did not consider here. However, in reality one never actually measures the particle position with infinite precision; one always has a finite grid of positions. Therefore, while it may be mathematically more convenient to use an infinite-dimensional Hilbert-space, the system can in principle be described using a finite Hilbert space.}, for Stern-Gerlach, and for {\sc EPR}-type
experiments, no matter how the settings were chosen. In addition, the dynamical law that we have made up here is both deterministic and local: It's not like the prepared state actually interacts with the detector before measurement, it just has the required information already in the dynamical law. We have found a local, deterministic, hidden variable model that reproduces quantum mechanics on the average without a hint of conspiracy. 

Now, for what the economy of this model is concerned, one may debate whether replacing the collapse postulate with random variables is a good trade-off. It certainly is a conceptual advantage, because it gives a well-defined probabilistic interpretation to quantum mechanics. However, if we merely count axioms, our toy model has no advantage over quantum mechanics. This is not surprising: If one believes in ``shut up and calculate'' and a probabilistic prediction is all one wants, then one can as well stick with quantum mechanics. But the here discussed toy model demonstrates that there is nothing per se wrong with theories that violate measurement independence, and that they are hence a viable route towards a theory more fundamental than quantum mechanics.

\section{Conclusion}

 We have shown that fine-tuning, which has frequently been used as an argument against superdetermism, it unnecessary: To make predictions with this model one does not need to introduce a huge number of delicately chosen parameters. This toy model shows that  violating statistical independence is a scientifically sound option for resolving the quantum measurement problem. 

The model, when averaged over the uniformly distributed hidden variables $\lambda$, reproduces the quantum mechanical predictions. If the variables were not uniformly distributed, for example because the sample space is not large enough, this model would give predictions that differ from quantum mechanics. That is, even in an experiment where one expects to have a good enough sampling to reproduce Born's rule with only small statistical deviations, the observed distribution could still be heavily skewed in the space of hidden variables, hence not approximate Born's rule as expected. This makes clear that superdeterminism is not an interpretation of quantum mechanics; it is a concrete possibility for replacing quantum mechanics with a better theory and more effort should be made to experimentally test it.

For example, following \cite{Hossenfelder:2011ct}, consider the case in which the values of the hidden variables are determined by the degrees of freedom of the measurement device (other than the measurement setting itself). Then, if we perform multiple measurements of non-commuting observables using the same setup, but the degrees of freedoms of the device which determine the $\lambda$s only change very slowly between the measurements, we might end up sampling always the same region of the parameter space of the $\lambda$s, hence observing results which are more strongly correlated than quantum mechanics predicts.

Exactly when the deviations from quantum mechanics become non-negligible depends on what the hidden variables are; the above introduced toy model cannot answer this question. The toy model is therefore, strictly speaking, untestable, because it does not specify where the distribution of hidden variables comes from. But, as pointed out above and in more detail in \cite{Hossenfelder:2011ct,Hossenfelder:2019shy}, if the hidden variables are the degrees of freedom of the detector, it is reasonable to expect that
minimizing the variation in the detectors' degrees of freedom between consecutive measurements will reveal deviations from Born's rule which cannot be detected by Bell-type experiments \cite{Leung:2017ndn,bigbell,Friedman:2018byq}, no matter how ingeniously conceived and executed.

\section*{Acknowledgements}

We gratefully acknowledge support from the Fetzer Franklin Fund. SH thanks Tim Palmer for helpful discussion.

\clearpage

\section*{Appendix: General Solution for $N=4$}

As previously, we work in the limit where $\kappa$ is much larger than the typical energies of the system, hence in very good approximation we can set $H=0$ . We begin with rewriting Eq.\ (\ref{dynN}) more explicitly inserting Eq. (\ref{lnk}):
\beqn
\partial_{t}\rho(t) &=& 
\kappa\sum_{M>K=1}^{4}\left(L_{MK}^{\textcolor{white}{\dagger}}\rho(t)L_{MK}^{\dagger}-\frac{1}{2}\{\rho(t),L_{MK}^{\dagger}L^{\textcolor{white}{\dagger}}_{MK}\}\right) \nonumber \\
&=&
\kappa\sum_{M>K=1}^{4}\left[\theta(\sigma_{M})\left(|M\rangle\langle K|\rho(t)|K\rangle\langle M|-\frac{1}{2}\{\rho(t),|K\rangle\langle K|\}\right) \right. \nonumber \\
&+& \left. \theta(-\sigma_{M})\left(|K\rangle\langle M|\rho(t)|M\rangle\langle K|-\frac{1}{2}\{\rho(t),|M\rangle\langle M|\}\right)\right]~.
\eeqn
We now introduce the notation $\rho_{IJ}(t):=\langle I|\rho(t)|J\rangle$ by use of which the master-equation takes the form
\beqn\label{gen}
\partial_{t}\rho_{IJ}(t) &=& 
\kappa\sum_{M>K=1}^{4}\left[
\theta(\sigma_{M})\left(\delta_{IM}\delta_{MJ}\rho_{KK}(t)-\frac{1}{2}\delta_{KJ}\rho_{IK}(t)-\frac{1}{2}\delta_{IK}\rho_{KJ}(t)\right)\right. \nonumber \\
 &+& \left.  \theta(-\sigma_{M})\left(\delta_{IK}\delta_{KJ}\rho_{MM}(t)-\frac{1}{2}\rho_{IM}(t)\delta_{MJ}-\frac{1}{2}\delta_{IM}\rho_{MJ}(t)\right) \right]~.
\eeqn
To apply the Kronecker delta, we recall
that the double sum can be equivalently written as:
\begin{equation}
\sum_{M>K}^{4}=\sum_{M=2}^{4}\sum_{K=1}^{M-1}=\sum_{K=1}^{3}\sum_{M=K+1}^{4}~.
\end{equation}
Using this, we obtain for $I=J$
\beqn
 \partial_{t}\rho_{JJ}(t)/\kappa &=&
 \sum_{M=2}^{4}\sum_{K=1}^{M-1}\theta(\sigma_{M})\delta_{JM}\delta_{MJ}\rho_{KK}(t)-\frac{1}{2}\sum_{K=1}^{3}\sum_{M=K+1}^{4}\theta(\sigma_{M})\delta_{KJ}\rho_{JK}(t)\nonumber \\
&-&\frac{1}{2}\sum_{K=1}^{3}\sum_{M=K+1}^{4}\theta(\sigma_{M})\delta_{JK}\rho_{KJ}(t) +
 \sum_{K=1}^{3}\sum_{M=K+1}^{4}\theta(-\sigma_{M})\delta_{JK}\delta_{KJ}\rho_{MM}(t) \nonumber \\
&-&\frac{1}{2}\sum_{M=2}^{4}\sum_{K=1}^{M-1}\theta(-\sigma_{M})\rho_{JM}(t)\delta_{MJ}-
\frac{1}{2}\sum_{M=2}^{4}\sum_{K=1}^{M-1}\theta(-\sigma_{M})\delta_{JM}\rho_{MJ}(t) \nonumber \\
&=& (1-\delta_{J1})\theta(\sigma_{J})\sum_{K=1}^{J-1}\rho_{KK}(t)
-\frac{1}{2}(1-\delta_{J4})\rho_{JJ}(t)\sum_{M=J+1}^{4}\theta(\sigma_{M}) \nonumber \\
&-&\frac{1}{2}(1-\delta_{J4})\rho_{JJ}(t)\sum_{M=J+1}^{4}\theta(\sigma_{M}) 
+
(1-\delta_{J4})\sum_{M=J+1}^{4}\theta(-\sigma_{M})\rho_{MM}(t)\nonumber \\
&-&
\frac{1}{2}(1-\delta_{J1})\sum_{K=1}^{J-1}\theta(-\sigma_{J})\rho_{JJ}(t)-\frac{1}{2}(1-\delta_{J1})\sum_{K=1}^{J-1}\theta(-\sigma_{J})\rho_{JJ}(t) \nonumber \\
&=&
(1-\delta_{J1})\left(\theta(\sigma_{J})\sum_{K=1}^{J-1}\rho_{KK}(t)-\sum_{K=1}^{J-1}\theta(-\sigma_{J})\rho_{JJ}(t)\right) \nonumber \\
&+&(1-\delta_{J4})\left(\sum_{M=J+1}^{4}\theta(-\sigma_{M})\rho_{MM}(t)-\rho_{JJ}(t)\sum_{M=J+1}^{4}\theta(\sigma_{M})\right) \nonumber \\
&=&
(1-\delta_{J1})\left[\theta(\sigma_{J})\sum_{K=1}^{J-1}\rho_{KK}(t)-\theta(-\sigma_{J})\rho_{JJ}(t)(J-1)\right] \nonumber \\
&+& (1-\delta_{J4})\sum_{M=J+1}^{4}\left[\theta(-\sigma_{M})\rho_{MM}(t)-\theta(\sigma_{M})\rho_{JJ}(t)\right]~.
\eeqn
This can be written explicitly as
\beqn
\partial_{t}\rho_{11}(t)/\kappa &=& \theta(-\sigma_{2})\rho_{22}(t)+\theta(-\sigma_{3})\rho_{33}(t)+\theta(-\sigma_{4})\rho_{44}(t) \nonumber \\
&-&\sum_{M=2}^{4}\theta(\sigma_{M})\rho_{11}(t)~,  \label{rho11} \\
\partial_{t}\rho_{22}(t)/\kappa &=&
 \theta(\sigma_{2})\rho_{11}(t)-\left(\theta(-\sigma_{2})+\theta(\sigma_{3})+\theta(\sigma_{4})\right)\rho_{22}(t) \nonumber \\
&+&\theta(-\sigma_{3})\rho_{33}(t)+\theta(-\sigma_{4})\rho_{44}(t)~,  \\
\partial_{t}\rho_{33}(t)/\kappa &=& \theta(\sigma_{3})\rho_{11}(t)+\theta(\sigma_{3})\rho_{22}(t) \nonumber \\ &-&\left(2\theta(-\sigma_{3})+\theta(\sigma_{4})\right)\rho_{33}(t)+\theta(-\sigma_{4})\rho_{44}(t)~, \\
\partial_{t}\rho_{44}(t)/\kappa  &=& \theta(\sigma_{4})\rho_{11}(t)+\theta(\sigma_{4})\rho_{22}(t)+\theta(\sigma_{4})\rho_{33}(t)-3\theta(-\sigma_{4})\rho_{44}(t)~. \label{rho44}
\eeqn

For $I\neq J$ Eq. (\ref{gen}) becomes:
\beqn
\partial_{t}\rho_{IJ}(t)/\kappa &=& \sum_{K=1}^{3}\sum_{M=K+1}^{4}\theta(\sigma_{M})\left(-\frac{1}{2}\delta_{KJ}\rho_{IK}(t)-\frac{1}{2}\delta_{IK}\rho_{KJ}(t)\right) \nonumber \\
&+&\sum_{M=2}^{4}\sum_{K=1}^{M-1}\theta(-\sigma_{M})\left(-\frac{1}{2}\rho_{IM}(t)\delta_{MJ}-\frac{1}{2}\delta_{IM}\rho_{MJ}(t)\right) \nonumber \\
&=&-\frac{1}{2}\left(\sum_{M=J+1}^{4}\theta(\sigma_{M})(1-\delta_{J4})\rho_{IJ}(t)+\sum_{M=I+1}^{4}(1-\delta_{I4})\theta(\sigma_{M})\rho_{IJ}(t)\right)\nonumber \\
&-&\frac{1}{2}\left((1-\delta_{J1})\sum_{K=1}^{J-1}\theta(-\sigma_{J})\rho_{IJ}(t)+(1-\delta_{I1})\sum_{K=1}^{I-1}\theta(-\sigma_{I})\rho_{IJ}(t)\right)\nonumber \\
&=&-\frac{1}{2}\left[\left((1-\delta_{J4})\sum_{M=J+1}^{4}\theta(\sigma_{M})+(1-\delta_{I4})\sum_{M=I+1}^{4}\theta(\sigma_{M})\right) \right. \nonumber \\
&+& \left. \left((1-\delta_{J1})\sum_{K=1}^{J-1}\theta(-\sigma_{J})+(1-\delta_{I1})\sum_{K=1}^{I-1}\theta(-\sigma_{I})\right)\right]\rho_{IJ}(t)\nonumber \\
&=&-\frac{1}{2}\left[\left((1-\delta_{J4})\sum_{M=J+1}^{4}\theta(\sigma_{M})+(1-\delta_{I4})\sum_{M=I+1}^{4}\theta(\sigma_{M})\right) \right. \nonumber \\
&+&\left. \left((1-\delta_{J1})(J-1)\theta(-\sigma_{J})+(1-\delta_{I1})(I-1)\theta(-\sigma_{I})\right)\right]\rho_{IJ}(t)~.
\eeqn
With that, we can write 
for $I\neq J$
\begin{equation}
\partial_{t}\rho_{IJ}(t)=-\frac{\kappa}{2}\Lambda_{IJ}\rho_{IJ}(t)~,\label{I neq J}
\end{equation}
where
\beqn
\Lambda_{IJ} &:=&\left((1-\delta_{J4})\sum_{M=J+1}^{4}\theta(\sigma_{M})+(1-\delta_{I,4})\sum_{M=I+1}^{4}\theta(\sigma_{M})\right) \nonumber \\
&+&\left((1-\delta_{J1})(J-1)\theta(-\sigma_{J})+(1-\delta_{I1})(I-1)\theta(-\sigma_{I})\right)~.
\eeqn
Eq. (\ref{I neq J}) can be integrated straight-forwardly to
\begin{equation}
\rho_{IJ}(t)=\rho_{IJ}(0)\,e^{-\frac{\kappa}{2}\Lambda_{IJ}t}.\label{Sol I neq J}
\end{equation}

For the singlet state we always have $\lambda_{3,4}\le0$
so Eqs.\ (\ref{rho11}) - (\ref{rho44}) simplify to
\beqn
\partial_{t}\rho_{11}(t) &=&\kappa\left[\theta(-\sigma_{2})\rho_{22}(t)+\rho_{33}(t)+\rho_{44}(t)-\theta(\sigma_{2})\rho_{11}(t)\right]~, \\
\partial_{t}\rho_{22}(t) &=&\kappa\left[\theta(\sigma_{2})\rho_{11}(t)-\theta(-\sigma_{2})\rho_{22}(t)+\rho_{33}(t)+\rho_{44}(t)\right]~,\\
\partial_{t}\rho_{33}(t) &=& -2\kappa\rho_{33}(t)+\kappa\rho_{44}(t)~,\\
\partial_{t}\rho_{44}(t) &=& -3\kappa\rho_{44}(t)~.
\eeqn
We solve these equations starting from the last one 
\beqn
\rho_{44}(t) &=& \rho_{44}(0)e^{-3\kappa t}\label{ro_44 sol}~.
\eeqn
This can be inserted into the third equation and gives
\beqn
\rho_{33}(t) &=& -\rho_{44}(0)e^{-3\kappa t}+\left(\rho_{33}(0)+\rho_{44}(0)\right)e^{-2\kappa t}.\label{ro_33 sol}
\eeqn
Inserting the initial state (\ref{rhoinitial}) now reveals that the solution to Eqs.\ (\ref{ro_44 sol})
and (\ref{ro_33 sol}) is simply $\rho_{33}(t)=\rho_{44}(t)=0$.

Then the first two equations are
\beqn
\partial_{t}\rho_{11}(t) &=&\kappa\theta(-\sigma_{2})\rho_{22}(t)-\kappa\theta(\sigma_{2})\rho_{11}(t)~,\\
\partial_{t}\rho_{22}(t) &=&\kappa\theta(\sigma_{2})\rho_{11}(t)-\kappa\theta(-\sigma_{2})\rho_{22}(t)~,
\eeqn
which can be integrated to
\beqn
\rho_{11}(t) &=& \theta(\sigma_2) \rho_{11}(0)e^{-\kappa t} + \theta(-\sigma_2) \left( \rho_{11}(0)+\rho_{22}(0)\left(1-e^{-\kappa t}\right) \right)~,\\
\rho_{22}(t) &=& \theta(\sigma_2) \left( \rho_{22}(0)+\rho_{11}(0)\left(1-e^{-\kappa t}\right) \right) + \theta(-\sigma_2) \rho_{22}(0)e^{-\kappa t} ~.
\eeqn
After inserting the initial state $\rho_{11}(0)=q^{2}$ and $\rho_{22}(0)=1-q^{2}$
we get for the diagonal elements
\beqn
\rho_{11}(t) &=&\theta(\sigma_{2})q^{2}e^{-\kappa t}+\theta(-\sigma_{2})\left[1-(1-q^{2})e^{-\kappa t}\right]\label{rho11 final}~,\\
\rho_{22}(t) &=&\theta(\sigma_{2})\left(1-q^{2}e^{-\kappa t}\right)+ \theta(-\sigma_{2})\left(1-q^{2}\right)e^{-\kappa t}\label{rho22 final-1}~.
\eeqn
For the off-diagonal elements of the density matrix, we note that all contributions to $\Lambda_{IJ}$ for $I,J$ equal to $3$ or $4$ vanish and we only have one non-zero matrix element which is
\begin{equation}
\Lambda_{12}=\theta(\sigma_{2})+\theta(-\sigma_{2})=1~.
\end{equation}
then
\begin{equation}
\rho_{12}(t)=-q\sqrt{1-q^{2}}\,e^{-\frac{\kappa}{2}t}\label{Sol I neq J-1}~.
\end{equation}
Eqs.\ (\ref{rho11 final}), (\ref{rho22 final-1}), and (\ref{Sol I neq J-1}) are used in (\ref{solsN4}).

\end{document}